\documentstyle[multicol,aps,prb,epsfig]{revtex}
\begin{document}

\title{Numerical verification of universality for
the Anderson transition}

\author{Keith Slevin}
\address{Department of Physics, Graduate School of Science,
Osaka University, \\ 1-1 Machikaneyama, Toyonaka,
Osaka 560-0043, Japan}

\author{Tomi Ohtsuki}
\address{Department of Physics, Sophia University,
Kioi-cho 7-1, Chiyoda-ku, Tokyo 102-8554, Japan}

\date{Journal Ref: Physical Review B {\bf 63} 45108 (2001)}

\maketitle

\begin{abstract}
We analyze the scaling behavior of the higher Lyapunov exponents
at the Anderson transition. 
We estimate the critical exponent and verify its universality and
that of the critical conductance distribution for box, Gaussian and
Lorentzian distributions of the random potential.
\end{abstract}

\begin{multicols}{2}

\section{Introduction}

Universality is a concept that is central to the theory of 
critical phenomena and continuous phase transitions.
The idea is that critical phenomena are independent
of the ``details'' of the physical system depending
only on its symmetry and dimensionality.
All systems in the same ``universality class'' should 
exhibit the same critical phenomena \cite{gold}.

The Anderson transition is a continuous phase transition 
and so the concept of universality should apply.
In this case, the important symmetries are those of time reversal
and spin rotation.
Three universality classes are expected: orthogonal, unitary, and
symplectic.
In this paper we are concerned with
the orthogonal universality class, which is comprised of
systems that have both time reversal and spin rotation symmetry.
This corresponds to an electron in a random potential in the absence 
of a magnetic field and with negligible spin-orbit interaction
\cite{LR,KM}.

One of the simplest ways to test the validity of the concept of
universality is to change the distribution of the random potential
in which the electron moves.
Since this does not change the universality class, the critical exponents
for different distributions of the random potential 
should be the same \cite{BSK}.
In a recent paper \cite{SO1} we presented numerical estimates of the critical
exponent for the Anderson transition for three different distributions
of the random potential: box, Gaussian and Lorentzian.
The estimates were made by analyzing the finite size scaling of the
localization length of electrons on long wires, and the results
support the contention that the critical exponent has a universal value 
independent of the distribution of the random potential.

In this paper, we analyze the scaling of the higher Lyapunov exponents
\cite{markos2} for the box, Gaussian and Lorentzian distributions of 
the random potential.
We give a clear demonstration of single parameter scaling for these 
higher Lyapunovs,  a clear verification of the universality of this
scaling with respect to these three distributions of random potential,
and show that this scaling yields estimates of the 
critical exponent that are in almost exact agreement with our previous
estimates based on the scaling of only the lowest Lyapunov exponent 
\cite{SO1}. (The lowest Lyapunov exponent is the inverse of the 
localization length.)

We also present an estimation of the critical conductance distribution 
$p_c(g)$. This is expected to take a size independent universal form at 
the critical point \cite{shapiro}. The size independence of $p_c(g)$ has 
been verified previously \cite{markos,SO2,SO3}. Here, we verify that the form 
of $p_c(g)$ is independent of the distribution of the random potential for 
the three distributions mentioned.

\section{Scaling of the higher Lyapunov exponents}

\subsection{Model and method}

The motion of the electrons is described by the Anderson model
\begin{equation}
 H = V \sum_{<i,j>} C_i^{\dagger}C_j +
     \sum_i W_i C_i^{\dagger}C_i ,
\end{equation}
where $C_i^{\dagger}(C_i)$ denotes the creation (annihilation)
operator of an electron at the site $i$ of a three dimensional
cubic lattice.
The value of the potential at site $i$ is $W_i$.
Hopping is restricted to nearest neighbors and its
amplitude is taken as the unit of energy, $V=1$.

We have examined three different distributions of the random potential.
\begin{enumerate}
	\item A box distribution with each $W_i$ uniformly distributed on the
	interval $[-W/2,W/2].$
	\item A Gaussian distribution with mean zero and variance $W^2/12$.
	\item A Lorentzian distribution of the form
	\begin{equation}
	p(W_i) = \frac{W}{\pi \left(W_i^2 + W^2\right)},
	\end{equation}
	for which the variance diverges.
\end{enumerate}
The standard method used in transfer matrix studies of the Anderson
transition \cite{MK} involves rewriting the Schr\"oedinger equation for electrons
on a long bar with dimensions $L\times L \times L_z$ as a product
of transfer matrices
\begin{equation}
M = \prod_{n=1}^{L_z} M_n,
\label{rmp}
\end{equation}
where $M_n$ is the transfer matrix. From $M$ we can define a matrix $\Omega$ by
\begin{equation}
\Omega = \ln M M^{\dagger},
\end{equation}
with eigenvalues $\{-\nu_N,\dots,-\nu_1,+\nu_1,\dots,+\nu_N \}$
where $N=L^2$ and the $\nu's$ are arranged in increasing order.
The eigenvalues occur in pairs of opposite sign as
a consequence of current conservation \cite{SN1,SN2} and
therefore we need only consider the positive $\nu's$.
The Lyapunov exponents $\{ \alpha_n \}$ for the random matrix product 
(\ref{rmp}) can be defined as
\begin{equation}
\alpha_n = \lim_{L_z \rightarrow \infty} = \frac{\nu_n}{2 L_z}.
\end{equation}
These were calculated to within a specified accuracy using the method
described in \cite{MK}.
The localization length $\lambda$ for electrons on the bar is the reciprocal 
of the smallest positive Lyapunov exponent
$\lambda = 1/\alpha_1$.
In our previous work, we examined only the scaling of the quantity
$\Lambda = \lambda/L=1/\alpha_1 L$.
Here we have examined the scaling of the quantities $z_n$ defined by
\begin{equation}
z_n = \alpha_n L,
\end{equation}
for higher values of $n$.
In performing the scaling analysis corrections to scaling were 
taken into account using the method described in (Ref. 5). 
We assumed that the data obey the following scaling law
\begin{equation}
z_n = F^{(n)} \left( \psi L^{1/\nu} , \phi L^y \right),
\end{equation}
where $\psi$ is a relevant scaling variable and $\phi$ is an 
irrelevant scaling variable.
We approximated this scaling function by its first order expansion
in the irrelevant scaling variable and fitted the data
for each $z_n$ to the form
\begin{equation}
z_n = F_0^{(n)}(\psi L^{1/\nu}) + \phi L^y F_1^{(n)}(\psi L^{1/\nu}).
\end{equation}
The scaling variables were approximated by expansions
in terms of the dimensionless disorder $w=(W_c-W)/W_c$ where
$W_c$ is the critical
disorder separating the insulating and metallic phases,
\begin{equation}
\psi = \psi_1 w + \psi_2 w^2, \ \ \ \ \ \phi=\phi_0.
\end{equation}
The critical exponent $\nu$ describes the divergence of the 
localization length (correlation length on the metallic side)
as the transition is approached
\begin{equation}
\xi = \xi_0 \left| \psi \right|^{-\nu}.
\end{equation}
Note that the absolute scale of the localization length $\xi_0$ 
cannot be determined with the method we have used here.
The decay of the irrelevant scaling variable with system size
is described by an exponent $y<0$.

The functions
$F_0^{(n)}$ and $F_1^{(n)}$ were expanded either to second or third order in $w$.
Also, when possible, the expansion for $\psi$ was truncated at 
the first order in $w$.
The decision at which order to truncate these various series was based
on the criteria of obtaining an acceptable goodness of fit with as few
parameters as possible.

\subsection{Results}

We simulated systems with sizes $L=4,5,6,8,10,12,14$ and $16$ and 
Fermi energy $E_F=0$ calculating data to $0.1\%$ accuracy with
the exception of some data very close to the critical point 
that were calculated to $0.05\%$ accuracy.
We found that the $z_n$'s obey single parameter scaling 
provided that the system sizes considered are large enough.
The minimum system size required increased with $n$ so that, 
for example, we could fit only the data with $L\ge 10$ when
analyzing $z_{10}$ for the box distribution.
In practice, we could demonstrate scaling for the first
ten exponents for the box and Gaussian distributions. 
For the Lorentzian distribution the required system sizes seem
to be larger and we could demonstrate scaling only for the
first three exponents.

\begin{figure}
\center
\epsfig{figure=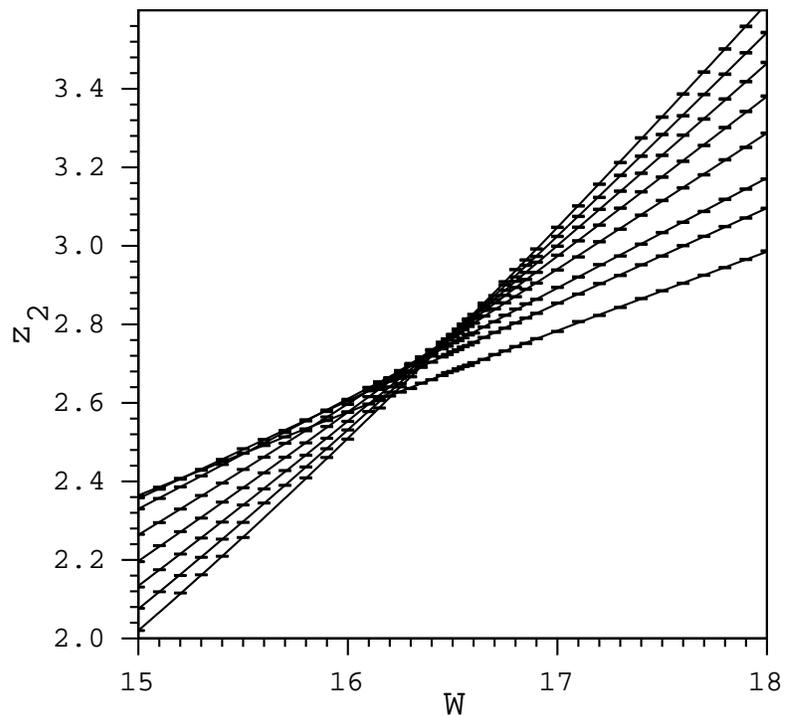,width=0.6\linewidth}
\caption{$z_2$ vs disorder $W$ for the box distributed random potential.}
\label{F1}
\end{figure}

\begin{figure}[h]
\center
\epsfig{figure=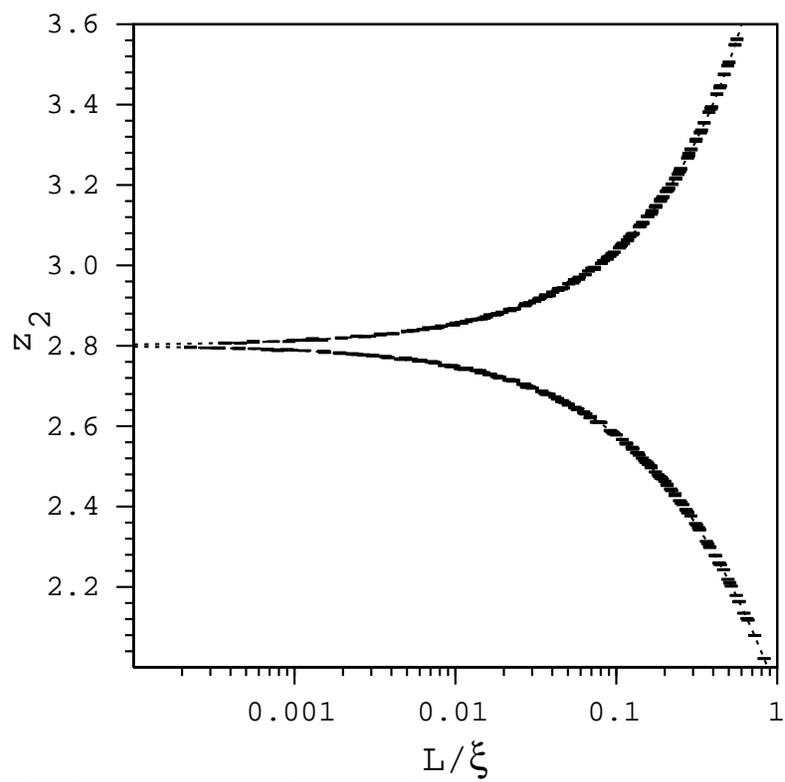,width=0.6\linewidth}
\caption{The same data as in Figure \ref{F1} after corrections to
scaling have been subtracted and plotted versus $L/\xi$ where $\xi$ is the
localization (correlation) length.}
\label{F2}
\end{figure}

\begin{figure}
\center
\epsfig{figure=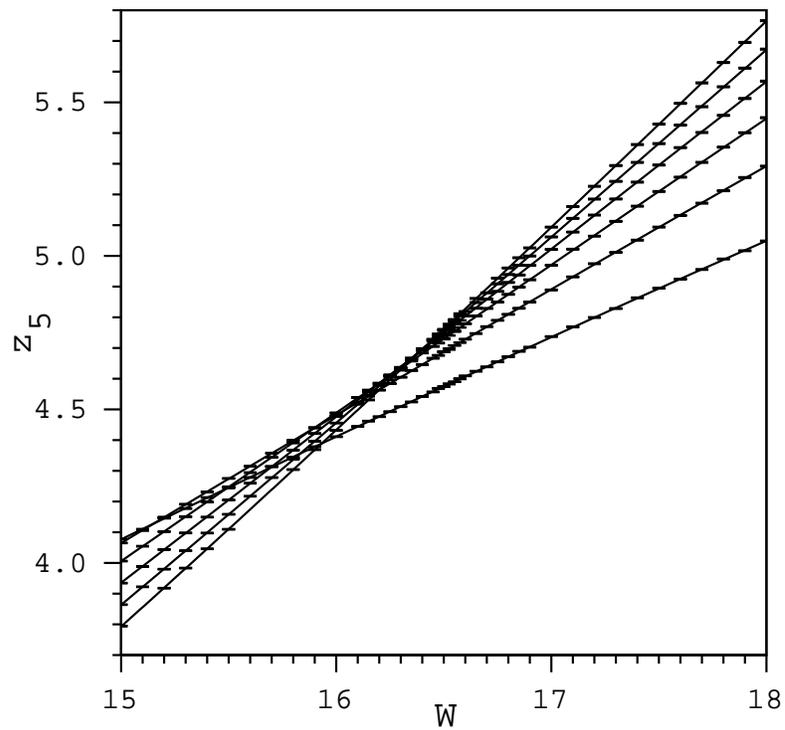,width=0.6\linewidth}
\caption{$z_5$ vs disorder $W$ for the Gaussian distributed 
random potential.}
\label{F3}
\end{figure}

\begin{figure}
\center
\epsfig{figure=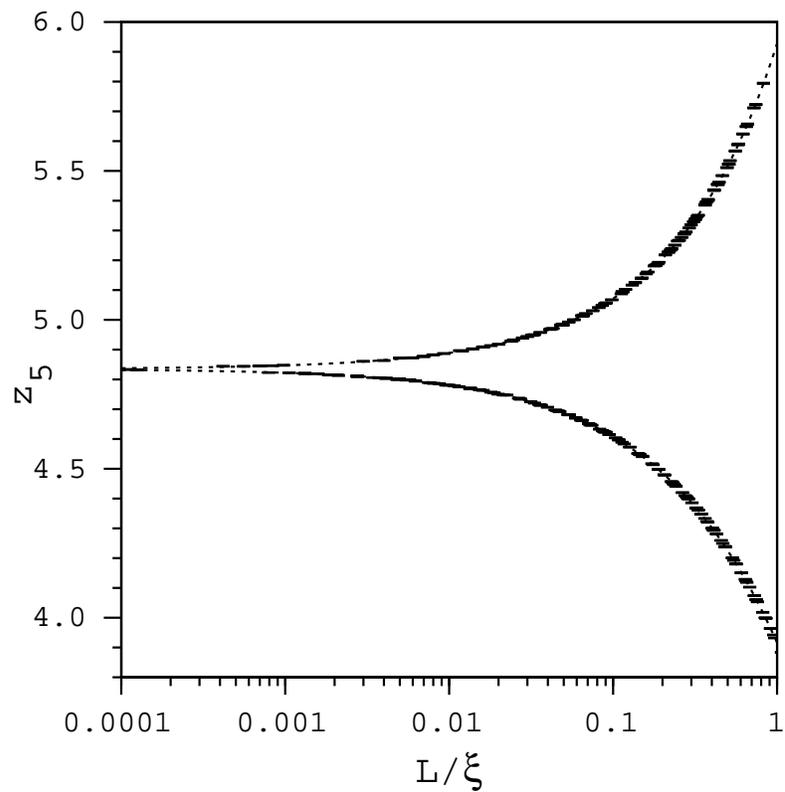,width=0.6\linewidth}
\caption{The same data as in Figure \ref{F3} after corrections to
scaling have been subtracted and plotted versus $L/\xi$.}
\label{F4}
\end{figure}

The results of the scaling analysis are presented in Tables \ref{T1}
-\ref{T3} and some of the data are plotted in Figs. 
\ref{F1}-\ref{F4}.
The estimates of the critical disorder obtained by scaling 
different Lyapunov exponents for a given distribution of the random
potential are in very close agreement.
The same can also be said for the critical values of the parameter
$z_n$ for various $n$ between different distributions of the
random potential. The critical values $z_c^{(n)}$ for a given $n$ is
defined by
\begin{equation}
z_c^{(n)} = F_0^{(n)}(0),
\label{eq10} 
\end{equation}
and is expected to be universal if the scaling function for
each Lyapunov is universal. The data clearly support this idea.
Turning to the critical exponent $\nu$, we find estimates that
are remarkably consistent across all the Lyapunov exponents 
and disorder distributions we considered.

\section{The zero temperature conductance distribution}

Next we considered the zero temperature conductance distribution
when $W=W_c$, i.e. the critical conductance distribution at the
Anderson transition.
We set the Fermi energy $E_F=0$ and, based on the results of 
Section II, we used $W_c=16.54$ for the box distribution, $W_c=21.3$ 
for the Gaussian distribution, and $W_c=4.26$ for the Lorentzian 
distribution when simulating the conductance distribution. 
An ensemble of samples of cubic shape was generated, and for each sample, a
Green's function iteration technique was used to determine the transmission
matrix $t$ in a two probe measuring geometry \cite{ando}. 
The dimensionless conductance $g$ was calculated using the Landauer formula 
\begin{equation}
g=2{\mathrm tr}tt^{\dagger},
\end{equation}
where the factor of two takes account of spin degeneracy.
(We would like to emphasize that the $t$ matrix that appears
in this formula is an $N\times N$ matrix, not an $L^2\times L^2$
matrix, where $N$ is the number of propagating modes
at the Fermi energy in the contacts attached at the left and 
right of the sample \cite{SN2}).

\begin{figure}
\center
\epsfig{figure=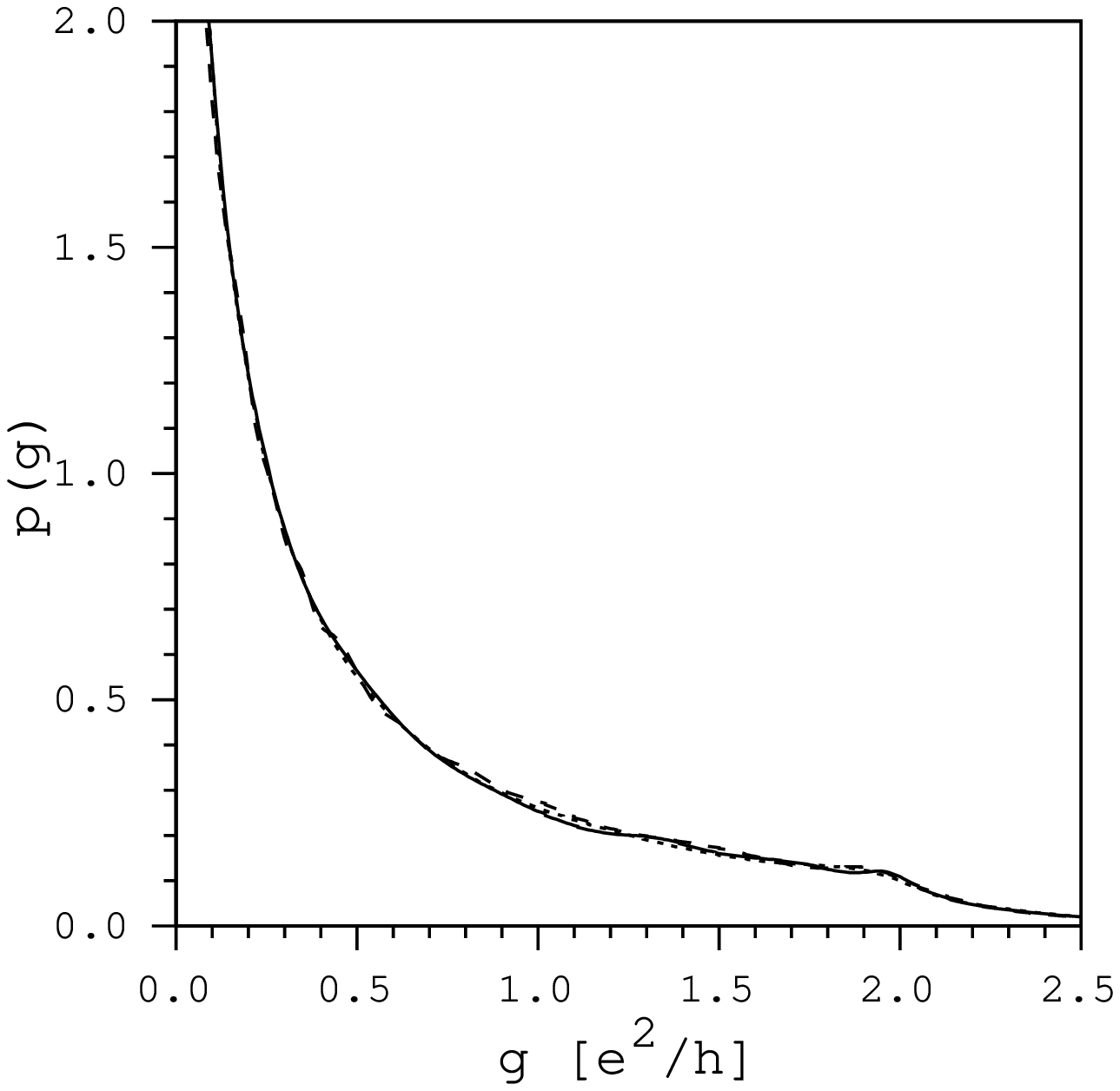,width=0.6\linewidth}
\caption{The distribution of the conductance for ensembles
of 100,000 $12 \times 12 \times 12$ cubes for box (solid line), Gaussian
(dotted line) and Lorentzian (dashed line) distributed random potentials.}
\label{F5}
\end{figure}

\begin{figure}
\center
\epsfig{figure=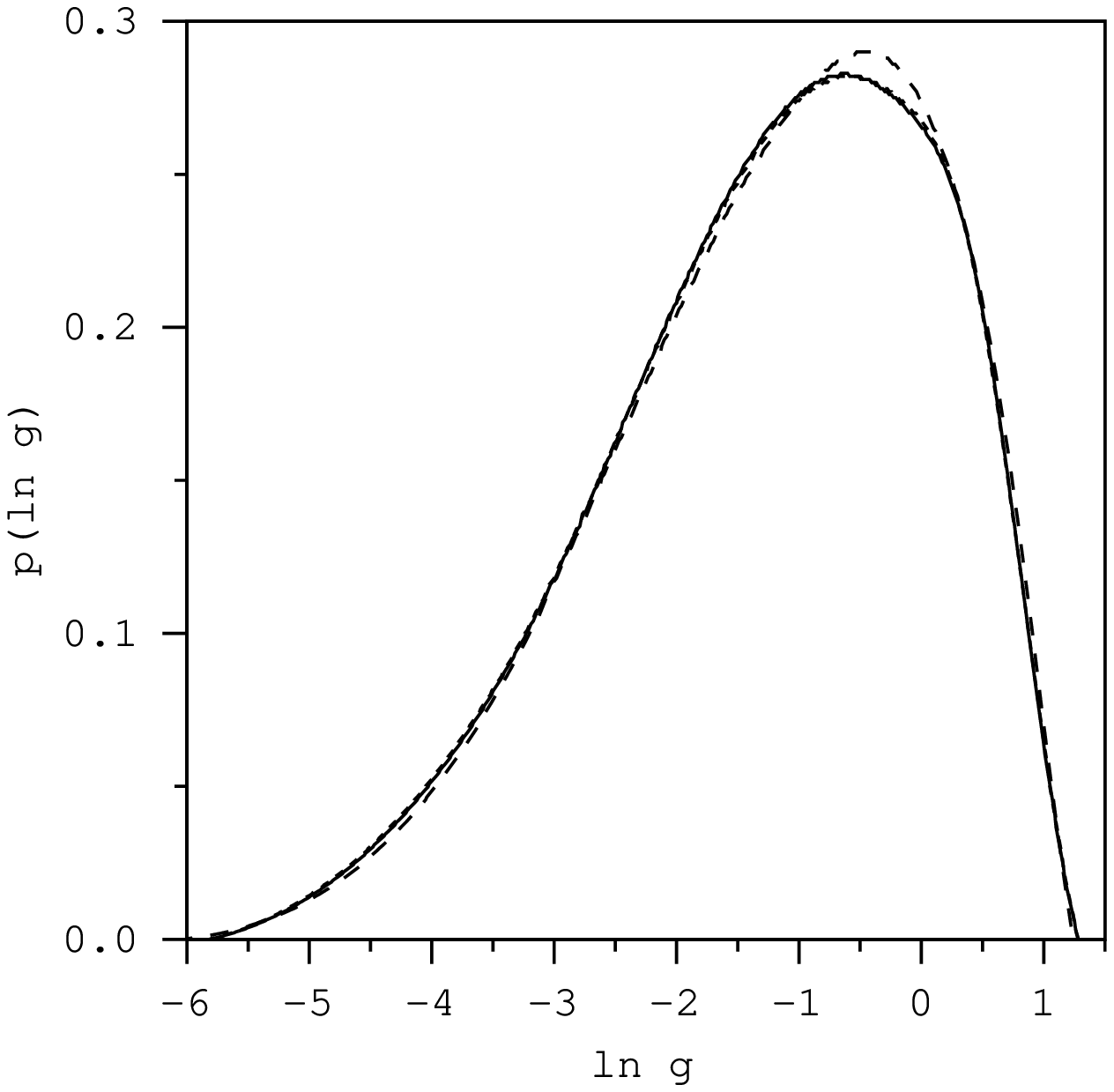,width=0.6\linewidth}
\caption{The distribution of the logarithm of the conductance
for ensembles of 100,000 $12 \times 12 \times 12$ cubes for box
(solid line), Gaussian (dotted line)
and Lorentzian (dashed line) distributed random potentials.}
\label{F6}
\end{figure}

In Fig. \ref{F5} and \ref{F6}, the conductance distribution for ensembles
of $12\times 12 \times 12$ cubic samples for the three distributions
of random potential are presented.
Fixed boundary conditions were imposed on the wavefunction
in the direction transverse to current flow so 
that the system size $L=12$ is large enough that the distribution 
shown are a good approximation
to the true critical distributions $p_c(g)$ and $p_c(\ln g)$
obtained in the asymptotic limit $L\rightarrow \infty$ \cite{SOK}. 
(Larger system sizes are required if periodic boundary conditions are 
considered and the asymptotic
form of the distribution is different \cite{SO3,SOK,soukoulis}.)
Note that despite the appearance of Fig. \ref{F5}, $p_c(g)
\rightarrow 0$ as $g \rightarrow 0$. This can be deduced from the
form of $p_c(\ln g)$ in Fig. \ref{F6} or a more detailed plot
of $p_c(g)$ near $g=0$ \cite{SOK,soukoulis}.

Looking at Figs. \ref{F5} and \ref{F6} we see that critical 
distributions obtained
for the three different distributions of the random potential are
almost identical. This is consistent with the assertion that the
critical conductance distribution is universal.

\section{Summary and Discussion}

The critical exponent for the Anderson model with random hopping
but no random potential has been estimated \cite{Cain} to be 
$\nu = 1.61\pm .07$. Since models with random hopping and random
potentials are in the same universality class, we should, and do, 
obtain estimates for the critical exponent that are consistent 
with Ref. 18. 

The computer time required to estimate the Lyapunov exponents
with the method used here \cite{MK} scales as $L^7$ for a 
given accuracy.
Thus, it is unlikely that this method can be extended to
much larger systems sizes in the near future.
The critical exponent of the Anderson transition has also been 
estimated by analyzing the scaling of energy level statistics (ELS) \cite{shk}.
The most recent, and in our opinion the most accurate, estimate
of $\nu$ obtained with ELS is $\nu=1.45\pm0.2$ \cite{milde}. This is
consistent with our results but the estimated uncertainty is
an order of magnitude worse than we have achieved here.

In the simulations whose results we have presented here, the
Fermi energy is always set at the band center $E_F=0$.
For one dimensional systems it has been demonstrated 
that single parameter scaling does not hold for states with energies
in a band gap of the corresponding ordered
(i.e. zero disorder $W=0$) system \cite{Slevin,Deych1,Deych2} . In that case,
single parameter scaling is only observed for sufficiently
strong disorder.
We suspect that this might also hold for the higher Lyapunov exponents
in the three dimensional systems we have studied here for
energies outside the band.

To conclude, we have demonstrated that the higher Lyapunov exponents 
obey single 
parameter scaling at the Anderson transition and presented evidence
that this scaling behavior is universal and independent of the choice
of distribution of the random potential.
Analysis of this scaling behavior yields estimates of the critical 
exponent that are remarkably consistent between different Lyapunov
exponents and different distributions of the random potential.
Finally, we presented evidence that the critical conductance distribution
is also universal and independent of the choice of the distribution 
of the random potential.


\end{multicols}{2}

\begin{table}
\begin{tabular}{|l|l|l|l|l|l|}
$n$ & $N_d$ & $N_p$  & $W_c$ & $z_c^{(n)}$ & $\nu$ \\
1   & 336   & 12     & 16.54(53,55) & 1.737(733,742) & 1.56(55,58) \\
2   & 336   & 12     & 16.55(54,55) & 2.801(797,804) & 1.57(56,58) \\
3   & 292   & 12     & 16.55(55,56) & 3.596(591,602) & 1.57(56,58) \\
4   & 250   & 10     & 16.56(55,57) & 4.25(24,26)    & 1.58(57,59) \\
5   & 250   & 10     & 16.57(56,59) & 4.83(82,84)    & 1.58(57,59) \\
10  & 162   & 10     & 16.59(51,71) & 7.0(6.9,7.2)   & 1.55(52,59) \\
\end{tabular}
\caption{The results of the scaling analysis for the box 
distributed random potential. Here $n$ indicates which Lyapunov exponent
is being considered, $N_d$ the number of data available, $N_p$
the number of parameters used to fit the data, $W_c$ is the estimate
of the critical disorder, $z_c^{(n)}$ is defined in (\ref{eq10}) and 
$\nu$ is the estimated critical exponent. The figures in brackets 
are $95\%$ confidence intervals for the corresponding critical
parameter estimated using the bootstrap method (Ref. 14).}
\label{T1}
\end{table}

\begin{table}
\begin{tabular}{|l|l|l|l|l|l|}
$n$ & $N_d$ & $N_p$ & $W_c$ & $z_c^{(n)}$ & $\nu$ \\
1   & 200   & 10    & 21.30(28,31) & 1.737(734,742)  & 1.58(56,61) \\
2   & 200   & 9     & 21.29(28,30) & 2.796(794,799)  & 1.57(56,58) \\
3   & 175   & 9     & 21.31(30,32) & 3.593(588,597)  & 1.58(57,59) \\
4   & 150   & 9     & 21.31(29,33) & 4.244(236,252)  & 1.58(56,60) \\
5   & 125   & 9     & 21.29(27,33) & 4.80(79,82)     & 1.57(55,60) \\
10  & 100   & 9     & 21.40(30,57) & 7.05(6.98,7.18) & 1.55(50,62) \\
\end{tabular}
\caption{The results of the scaling analysis for the Gaussian 
distributed random potential. The notation is explained in the caption
for Table 1.}
\label{T2}
\end{table}

\begin{table}
\begin{tabular}{|l|l|l|l|l|l|}
$n$ & $N_d$ & $N_p$  & $W_c$ & $z_c^{(n)}$ & $\nu$ \\
1   & 256   & 10     & 4.25(24,26)  & 1.70(68,71) & 1.57(52,61) \\
2   & 192   & 10     & 4.24(23,26)  & 2.71(64,75)  & 1.58(53,65) \\
3   & 127   & 10     & 4.29         & 3.62         & 1.58        \\
\end{tabular}
\caption{The results of the scaling analysis for the Lorentzian 
distributed random potential. The notation is explained in the caption
for Table 1. (We could not execute the bootstrap method for $n=3$ so
no confidence intervals are available.)}
\label{T3}
\end{table}

\end{document}